\documentstyle{paper} 
\begin{document}
\newcommand{\Df}[2]{\mbox{$\frac{#1}{#2}$}}

\hfill  INP-96-11/418, hep-ph/9604254

\centerline{Explicit solutions of $n$-loop vacuum integral 
recurrence relations}

\centerline{\footnotesize
P.\,A.\,BAIKOV\footnote{
Supported in part
by the Russian Basic Research Foundation (grant N 96--01--00654),
INTAS (grant 91-1180).\\
\hspace*{3pt}Email: baikov@theory.npi.msu.su}}
\vspace*{0.015truein}
\centerline{\footnotesize\it Institute of Nuclear Physics,
Moscow State University}
\baselineskip=10pt   
\centerline{\footnotesize\it  119~899, Moscow, Russia}

\vspace{2mm}
\centerline{
\begin{minipage}{14cm}
{Explicit formulas for solutions of recurrence 
relations for 3--loop vacuum integrals are generalized for the $n$-loop case.}
\end{minipage}}

\vspace{1mm}

Recently \cite{hep} a new approach to implement recurrence 
relations \cite{ch-tk}  for vacuum integrals was suggested.
Such relations connect Feynman integrals with various
degrees of their denominators. In many cases
they provide a possibility to express an
integral with given degrees of denominators as a linear
combination of a few master integrals with some coefficient functions.
The common way to evaluate these functions is step--by--step recurrence 
procedure, which demands a lot of calculations.
On the other hand, the construction of such procedure is a serious problem
even at the three-loop level.

For vacuum three-loop integrals with one non-zero mass and various numbers of 
massless lines the corresponding algorithms were constructed in \cite{REC}.

In the approach proposed in \cite{hep} these coefficient functions were 
calculated directly
as solutions of the corresponding recurrence relations. 
In \cite{hep} explicit formulas for the solutions
of these relations for 3--loop case with arbitrary masses were obtained. 
As an example, the case of integrals with four equal masses and two massless 
lines was considered, and the efficiency
of this approach was demonstrated by calculations of the 3-loop QED vacuum 
polarization.

In this work we show how to extend the general formulas for the solutions
of the recurrence relations for the multi--loop case. 
Let us consider $L$-loop vacuum integrals with $N=L(L+1)/2$ denominators
(this number of denominators provides the possibility to represent any
scalar product of loop momenta as linear combination of the denominators;
the diagrams of practical interest which usually have less number of 
denominators, can be considered as partial cases when some degrees are 
equal to zero):

\begin{eqnarray}
B(\underline{n},D)\equiv
B(n_1,\ldots,n_N,D)=
\frac{m^{2\Sigma n_i-LD}}
{\big[\imath\pi^{D/2}\Gamma(3-D/2)\big]^L}
\int \cdots \int \frac{d^Dp_1\ldots d^Dp_L} 
{D_1^{n_1}\ldots D_N^{n_N}}
\label{integral}
\end{eqnarray}

\noindent
where $D_a=\sum_{i=1}^{N}\sum_{j=1}^{N}A^{(ij)}_a p_i\cdot p_j -\mu_a m^2$
and $p_i$ (${i=1,\ldots,L}$) are loop momenta. 

Let us derive recurrence relations that result from integration by parts,
by letting  $(\partial/\partial p_i)\cdot p_k$ act on
the integrand \cite{ch-tk}:

\begin{eqnarray}
D\delta_k^i B(\underline{n},D)&=&
2\sum_{a=1}^{N}\sum_{d=1}^{N}\sum_{l=1}^{L}A_d^{(il)} n_d{\bf I}^{d+} 
(A^{-1})_{(kl)}^a({\bf I}^-_a+\mu_a) B(\underline{n},D),
\label{rr}
\end{eqnarray}

\noindent
where  
${\bf I}^\pm_c B(n_1,\ldots, n_c,\ldots, )\equiv 
B(n_1,\ldots, n_c\pm1,\ldots, )$, and $(A^{-1})_{kl}^a$ come from the
expansion of the $p_i\cdot p_j$ terms appearing in the numerator through
denominators $D_a$: 

$$p_k\cdot p_l=\sum_{a=1}^N(A^{-1})_{(kl)}^a(D_a+\mu_a m^2).$$

Using the relations

$$[n_d{\bf I}^{d+},{\bf I}^-_a]=\delta_a^d,
\qquad
\sum_{a=1}^{N}A_a^{(il)} (A^{-1})_{(kj)}^a=
\frac{1}{2}(\delta^i_k\delta^l_j+\delta^i_j\delta^l_k)$$

\noindent
the recurrence relations (\ref{rr}) can be represented as

\begin{eqnarray}
\frac{D-L-1}{2}\delta_k^i B(\underline{n},D)&=&
\sum_{a=1}^{N}\sum_{d=1}^{N}\sum_{l=1}^{L}
(A^{-1})_{(kl)}^a({\bf I}^-_a+\mu_a) 
A_d^{(il)} n_d{\bf I}^{d+} 
B(\underline{n},D)
\label{rr1}
\end{eqnarray}

The differential equation corresponding to (\ref{rr1}) 
by substitutions 
$n_d{\bf I}^{d+} \rightarrow \partial/\partial x_d$,
${\bf I}^-_d \rightarrow x_d$ has a solution
$g(x_a)=P(x_a+\mu_a)^{(D-L-1)/2}$, where
$P(x_a)$ is the polynomial in $x_a$ of degree $L$:

$$P(x_a)=\det(\sum_{a=1}^N (A^{-1})_{(kl)}^a x_a)$$.

Let us consider "Laurent" coefficients of the function $g(x_a)$:

\begin{eqnarray}
f^k(\underline{n},D)=
\frac{1}{(2\pi\imath)^N}
\oint \cdots \oint
\frac
{dx_1 \cdots dx_N}
{x_1^{n_1} \cdots x_N^{n_N}}
\det((A^{-1})_{(kl)}^a(x_a+\mu_a))^{(D-L-1)/2}
\label{solution}
\end{eqnarray}

\noindent
where integral symbols denote $N$ subsequent complex 
integrations with contours 
which will be described below. Acting by (\ref{rr1}) on 
(\ref{solution}) one gets up to the surface terms 
the corresponding differential operator acting on 
$g(x_a)$, that is zero.
Then, the surface terms can be removed if we choose
closed or ended in infinity point contours (to provide the cancelation 
of the surface terms in the last case one can consider 
analytical continuations on $D$ from large negative values).
So (\ref{solution}) is a solution of relations (\ref{rr}),
and different choices of contours correspond to different 
solutions, enumerated by the index $k$. 

Then, according to dimensional regularization rules, integrals 
(\ref{integral}) are equal to zero if the number of positive $n_a$ is
less then $L$. 
In the  3--loop case \cite{hep} the coefficient
functions can be constructed as linear combinations
of those that are equal to zero if the index from
definite three--index set ("Taylor" indexes) is not positive. 
One can obtain such solutions 
if one chooses contours, corresponding to these indexes,
as circles around zero. In this case these three integrations can be
performed and lead to coefficient in the common Taylor expansion in
the corresponding variables.
We hope that it will be applicable as well for the $L$-loop case 
and the number 
of integrations in (\ref{solution}) can be reduced from $L(L+1)/2$  
to $L$.

Finally note that solutions (\ref{solution}) by construction satisfy 
D--shifted recurrence relations

\begin{eqnarray}
f^k(\underline{n},D)&=&P({\bf I}^-_a+\mu_a)f^k(\underline{n},D-2)
\label{rrD0}
\end{eqnarray}

\noindent
But if one want to construct the coefficient functions 
$\bar{f}^k(\underline{n},D)$ 
for some specific set of master integrals, one gets linear combinations
of $f^k(\underline{n},D)$ with coefficients depending on $D$.
This dependence leads to a nontrivial mixing matrix
in the D--shifted recurrence relations

\begin{eqnarray}
\bar{f}^k(\underline{n},D)&=
&\sum_n S^k_n(D)P({\bf I}^-_a+\mu_a)\bar{f}^n(\underline{n},D-2)
\label{rrD}
\end{eqnarray}

So in this paper we construct explicit formulas (\ref{solution})
for the solutions of the recurrence relations for multi--loop vacuum 
integrals with arbitrary masses. In forthcoming paper we plan to generalize
these formulas for the non-vacuum case.

\section*{Acknowledgment}

I would like to thank V.A.Smirnov for careful reading of this 
text.

\end{document}